\documentclass[10pt, twocolumn, final]{IEEEtran}
\usepackage{url}
\usepackage[dvips,final]{graphicx}
\usepackage{amssymb,amsfonts,amsmath,amsthm,amscd,dsfont,mathrsfs}
\usepackage{graphicx,float,psfrag,epsfig,color}
\usepackage{subfig}
\usepackage{nth}
\usepackage{mathtools}
\usepackage[noadjust]{cite}

\usepackage[numbers,sort&compress]{natbib}
\usepackage{balance}

\usepackage{mathabx}

\usepackage[none]{hyphenat}

\usepackage{mathtools}
\DeclarePairedDelimiter{\ceil}{\lceil}{\rceil}

\makeatletter
\def\maketag@@@#1{\hbox{\m@th\normalfont\normalsize#1}}
\makeatother

\newcommand{\subparagraph}{}
\usepackage[compact]{titlesec}
\titlespacing*{\section}{0pt}{0.8\baselineskip}{0.5\baselineskip}

\allowdisplaybreaks

\usepackage{tikz}
\usepackage{pgfplots}
\pgfplotsset{compat=newest}
\usetikzlibrary{patterns}
\usetikzlibrary{positioning}
\usetikzlibrary{datavisualization}
\usetikzlibrary{datavisualization.formats.functions}
\usetikzlibrary{backgrounds}
\usetikzlibrary{shapes,snakes}
\usepackage{tikz-cd}
\usetikzlibrary{matrix, calc, arrows}
\usepackage{adjustbox}

\usepgfplotslibrary{external} 
\usepackage{float}
\usepackage{afterpage}
\usepackage{balance}

\def\mindex#1{\index{#1}}

\def\sq{\hbox{\rlap{$\sqcap$}$\sqcup$}}
\def\qed{\ifmmode\sq\else{\unskip\nobreak\hfil
\penalty50\hskip1em\null\nobreak\hfil\sq
\parfillskip=0pt\finalhyphendemerits=0\endgraf}\fi\medskip}

\long\def\defbox#1{\framebox[.9\hsize][c]{\parbox{.85\hsize}{%
\parindent=0pt
\baselineskip=12pt plus .1pt      % STYLE
\parskip=6pt plus 1.5pt minus 1pt % CHANGES
 #1}}}

\long\def\beginbox#1\endbox{\subsection*{}%
\hbox{\hspace{.05\hsize}\defbox{\medskip#1\bigskip}}%
\subsection*{}}

\def\endbox{}

\def\rank{{\rm rank\,}}

\newsavebox{\junk}
\savebox{\junk}[1.6mm]{\hbox{$|\!|\!|$}}

\def\limsup{\mathop{\rm lim\ sup}}
\def\liminf{\mathop{\rm lim\ inf}}
\def\argmin{\mathop{\rm arg\, min}}

\def\argmax{\mathop{\rm arg\, max}}

\newcommand{\field}[1]{\mathbb{#1}}

\def\ind{\field{I}}

\def\bE{{\mathbb E}}
\def\bF{{\mathbb F}}

\def\bH{{\mathbb H}}
\def\bI{{\mathbb I}}

\def\bN{{\mathbb N}}

\def\bP{{\mathbb P}}

\def\bR{{\mathbb R}}

\def\bZ{{\mathbb Z}}

\def\bfC{{\bf C}}
\def\bfD{{\bf D}}

\def\bfG{{\bf G}}
\def\bfH{{\bf H}}

\def\bfU{{\bf U}}
\def\bfV{{\bf V}}

\def\bfX{{\bf X}}
\def\bfY{{\bf Y}}
\def\bfZ{{\bf Z}}

\def\bfw{{\bf w}}
\def\bfx{{\bf x}}
\def\bfy{{\bf y}}

\def\scrR{{\mathscr{R}}}

\def\sfA{{\sf A}}
\def\sfB{{\sf B}}

\def\sfH{{\sf H}}

\def\sfL{{\sf L}}

\def\sfR{{\sf R}}

\def\sfa{{\sf a}}
\def\sfb{{\sf b}}

\def\sfh{{\sf h}}

\def\sfr{{\sf r}}

\def\bfmath#1{{\mathchoice{\mbox{\boldmath$#1$}}%
{\mbox{\boldmath$#1$}}%
{\mbox{\boldmath$\scriptstyle#1$}}%
{\mbox{\boldmath$\scriptscriptstyle#1$}}}}

\def\bfmY{\bfmath{Y}}

\def\bfmhhaY{\bfmath{\hhaY}} %\widehat{\widehat{Y}}}}
\def\bfmhhaY{\hbox to 0pt{$\widehat{\bfmY}$\hss}\widehat{\phantom{\raise 1.25pt\hbox{$\bfmY$}}}}

\def\til={{\widetilde =}}

\def\clA{{\cal A}}

\def\clC{{\cal C}}
\def\clD{{\cal D}}
\def\clE{{\cal E}}
\def\clF{{\cal F}}

\def\clI{{\cal I}}

\def\clL{{\cal L}}

\def\clR{{\cal R}}
\def\clS{{\cal S}}
\def\clT{{\cal T}}

 \def\FRAC#1#2#3{\genfrac{}{}{}{#1}{#2}{#3}}

\def\ddtp{{\mathchoice{\FRAC{1}{d^{\hbox to 2pt{\rm\tiny +\hss}}}{dt}}%
{\FRAC{1}{d^{\hbox to 2pt{\rm\tiny +\hss}}}{dt}}%
{\FRAC{3}{d^{\hbox to 2pt{\rm\tiny +\hss}}}{dt}}%
{\FRAC{3}{d^{\hbox to 2pt{\rm\tiny +\hss}}}{dt}}}}

\def\average#1,#2,{{1\over #2} \sum_{#1}^{#2}}

\def\eye(#1){{\bf(#1)}\quad}

\newtheorem{theorem}{{\bf Theorem}}

\newtheorem{remark}{{\bf Remark}}
\newtheorem{definition}{{\bf Definition}}
\newtheorem{example}{{\bf Example}}
\newtheorem{conjecture}{{\bf Conjecture}}

\newtheorem{proposition}[theorem]{{\bf Proposition}}
\newtheorem{lemma}[theorem]{{\bf Lemma}}

\def\eq#1/{(\ref{e:#1})}

\newcommand{\beqn}[1]{\notes{#1}%
\begin{eqnarray} \elabel{#1}}

\newcommand{\eeqn}{\end{eqnarray} }

\newcommand{\beq}[1]{\notes{#1}%
\begin{equation}\elabel{#1}}

\newcommand{\eeq}{\end{equation}}

\def\bdes{\begin{description}}
\def\edes{\end{description}}

\newcounter{rmnum}

\newcounter{anum}

{\end{list}}

\def\ass(#1:#2){(#1\ref{#1:#2})}

\def\ritem#1{
\item[{\sf \ass(\current_model:#1)}]
}

\newenvironment{recall-ass}[1]{%
\begin{description}
\def\current_model{#1}}{
\end{description}
}

\long\def\comment#1{}

\newfont{\bbb}{msbm10 scaled 700}

\newfont{\bb}{msbm10 scaled 1100}

\newcommand{\Deltam}{\hbox{\boldmath$\Delta$}}

\newcommand{\transp}{{\sf T}}

\DeclarePairedDelimiter\quant{\langle}{\rangle}
\DeclarePairedDelimiter\floor{\lfloor}{\rfloor}
\DeclarePairedDelimiter\urid{\overline{d}(}{)}
\DeclarePairedDelimiter\lrid{\underline{d}(}{)}
\DeclarePairedDelimiter\mri{\mathsf{RI}(}{)}

\def\res{\mathsf{Res}}
\def\Thet{\Theta}

\def\contin{C}
\def\disc{D}

\newcommand{\lin}{{\cal L}}

\newcommand{\interv}[2]{{\text{$#1$:\,$#2$}}}

\begin{document}

\title{Polarization of the R\'enyi Information Dimension with Applications to Compressed Sensing}
\author{Saeid Haghighatshoar,  \IEEEmembership{Member, IEEE,} Emmanuel Abbe,
\IEEEmembership{Member, IEEE}\vspace{-1mm}
\thanks{This paper is the updated version of the paper \cite{HA2013polarization}, which was partially presented in ISIT 2013, Istanbul, Turkey \cite{HA2013polarization_isit}.}
\thanks{Saeid Haghighatshoar is with the Communications and Information Theory Group, Technische Universit\"{a}t Berlin (saeid.haghighatshoar@tu-berlin.de). Emmanuel Abbe has a joint position in Applied and Computational Mathematics and Electrical Engineering 
at Princeton University, New Jersey, USA (eabbe@princeton.edu). This work was started when both the authors were with the Information Processing Group (IPG), EPFL, Switzerland.}}

\maketitle

\begin{abstract}
In this paper, we show that the Hadamard matrix acts as an extractor over the reals of the R\'enyi information dimension (RID), in an analogous way to how it acts as an extractor of the discrete entropy over finite fields.   
More precisely, we prove that the RID of an i.i.d.\ sequence of mixture random variables polarizes to the extremal values of $0$ and $1$ (corresponding to discrete and continuous distributions) when transformed by a Hadamard matrix. Further, we prove that the polarization pattern of the RID admits a closed form expression and follows exactly the \textit{Binary Erasure Channel} (BEC) polarization pattern in the discrete setting. 
We also extend the results from the single- to the multi-terminal setting, obtaining a Slepian-Wolf counterpart of the RID polarization. 
We  discuss applications of the RID polarization to Compressed Sensing of i.i.d. sources. In particular, we use the RID polarization to construct a family of deterministic $\pm 1$-valued sensing matrices for Compressed Sensing. We run numerical simulations to compare the performance of the resulting matrices with that of  random Gaussian and random Hadamard matrices. The results indicate that the proposed matrices afford competitive performances while being explicitly constructed. 
\end{abstract}


\section{Introduction}
Let $X$ be a real-valued random variable. We denote the $q$-ary quantization of $X$ by 
$
\quant{X}_q= \frac{\floor{qX}}{q},
$
where for a real number $r$, we denote by $\floor{r}$ the largest integer less than or equal to $r$. The upper and the lower \textit{R\'enyi Information Dimension} (RID) of $X$ are defined by 
\begin{align}
\urid{X}&=\limsup _{q \to \infty} \frac{H(\quant{X}_q)}{\log_2(q)},\label{eq:urid}\\
\lrid{X}&=\liminf _{q \to \infty} \frac{H(\quant{X}_q)}{\log_2(q)},\label{eq:lrid}
\end{align}
where $H(\quant{X}_q)$ denotes the Shannon entropy of the discrete random variable $\quant{X}_q$ obtained from the  quantization. If the limits coincide, we define $d(X):=\urid{X}=\lrid{X}$. 
R\'enyi in his paper \cite{renyi} proved that if the random variable $X$ is discrete, continuos, or a mixture thereof,  the upper and the lower RID are equal, thus, $d(X)$ is well-defined. He also provided an example of a singular random variable for which these two limits do not coincide. Apart from being an information measure, the RID appears as the fundamental operational limit in diverse areas in probability theory and signal processing such as signal quantization \cite{gray2011entropy}, rate-distortion theory \cite{kawabata1994rate}, and fractal geometry \cite{falconer2004fractal}. More recently, the operational aspect of RID has reappeared in applications as varied as lossless analog compression \cite{WV1, alberti2016lossless, stotz2017almost}, Compressed Sensing of sparse signals \cite{montanari, li2012structured, discrete_partial_hadamard_isit}, and characterization of the degrees-of-freedom (DoFs) of vector interference channels \cite{wu2011degrees, stotz2012degrees}, which has recently been of significant importance in wireless  communication.

In this paper, motivated by \cite{renyi}, we first extend the definition of RID as an information measure from scalar random variables to a family of vector random variables over which the RID is well-defined. 
We also extend the definition to the joint and the conditional RIDs and provide a closed-form expression for computing them. 
Using these, we investigate the high-dimensional behavior of the RID of i.i.d. mixture random variables when transformed by a Hadamard matrix. 
We prove that the conditional RIDs of almost all the resulting random variables polarize to the extremal values of $0$ and $1$. We also obtain a formula for computing those conditional RIDs and their polarization pattern using the \textit{Binary Erasure Channel} (BEC) polarization in the discrete case \cite{polar_channel}. This gives a natural extension of the polarization phenomenon for the  entropy over the finite fields to the RID over the reals. 

We study some of the potential applications of the new polarization result in Compressed Sensing \cite{CS2, CS1, CS4, CS3}. 
In particular, motived by the recent results on the operational aspect of RID in Compressed Sensing \cite{WV1, montanari} and inspired by the success of polar codes in achieving information theoretic limits \cite{polar_source, polar_channel}, we exploit the RID polarization to design deterministic partial Hadamard matrices for Compressed Sensing of i.i.d. sparse signals. We compare the performance of the resulting matrices with that of other traditional matrices in Compressed Sensing such as random Gaussian and random Hadamard matrices.
Numerical simulations provide evidence that the constructed matrices together with recovery algorithms such as $l_1$-norm minimization provide a low-complexity Compressed Sensing and recovery procedure for the sparse signals. The use of polarization techniques for Compressed Sensing was also investigated independently in \cite{li2012structured}, approaching noiseless Compressed Sensing via a duality with analog channel coding. 

\subsection{Notation}\label{notation}
We use $\bR$ for the reals, $\bR_+$ for the positive reals, $\bZ$ for the integers, $\bZ_+$ for the set of positive integers, and $\bN$ for the set of strictly positive integers. We denote sets by calligraphic letters such as $\clA$ and their cardinality by $|\clA|$. We use capital letters for random variables and small letters for their realizations, e.g., $x$ is a realization of the random variable $X$. We denote the distribution of a random variable $X$ by  $p_X$. For $N=2^n$, we denote by $\bfH_N$ the standard Hadamard matrix of order $N$.  We use $[n]$ for the  set of integers $\{1,2, \dots, n\}$. We denote by $X_i^j$ the column vector $[X_i,X_{i+1}, \dots, X_j]^\transp$, where the vector is empty when $i>j$. Vectors are denoted by boldface letters, e.g., $\bfX=X_1^n$ and $\bfx=x_1^n$ denote an $n$-dim vector of random variables and its realization. For two sequences $f,h : \bN \to \bR$, we say $f_q \doteq h_q$ { if and only if }
\begin{align}
\lim_{q \to \infty} \frac{f_q-h_q}{\log_2(q)}=0.
\end{align}
We denote matrices by capital letters, e.g., $\sfA$. For an $m \times n$ matrix $\sfA$, and a subset of its columns $\clC \subseteq [n]$, we denote by $\sfA_\clC$ the $m\times |\clC|$ submatrix of $\sfA$ obtained by selecting those columns of $\sfA$ belonging to $\clC$. In a similar way, we denote by $\sfA[\clR]$ the $|\clR|\times n$ submatrix obtained by selecting the rows of $\sfA$ belonging to $\clR\subseteq [m]$. For two matrices $\sfA$ and $\sfB$, we denote by $[\sfA;\sfB]$  a matrix obtained by putting the rows of $\sfA$ on top of the rows of $\sfB$ and by $[\sfA,\sfB]$ the matrix obtained by putting the columns of $\sfB$ to the right of the columns of $\sfA$, provided that the resulting matrices are well-defined.

\subsection{Reminder on Polar Codes}
Polar codes were introduced by Arikan in his seminal paper \cite{polar_channel}. They are the first class of efficient codes that provably achieve channel capacity on all binary input symmetric channels. Recent research on polar codes has illustrated their theoretical optimality for other classical problems in information theory such as lossless and lossy source coding \cite{polar_source, polar_lossy_source}, coding over multiple-access channels (MAC) \cite{abbe_mac}, Wyner-Ziv and Gelfand-Pinsker problem \cite{polar_other}, and coding for secrecy over the wiretap channel \cite{mahdavifar2011achieving,sasoglu2013new}. 

The underlying structure behind all these applications of polar codes is the polarization phenomenon. To  explain briefly, we will mainly focus on the source coding aspect, which is more relevant to our work. Let  $\bfX=X_1^N,$ for $N=2^n$ a power of two, be a sequence of $N$ i.i.d. 
$\mathsf{Bernoulli}(p)$, $p \in (0,\frac{1}{2})$, random variables and let $\bfY= \bfG_N \bfX$, with the arithmetic over the binary field $\bF_2$, where $$\bfG_N=\left [ \begin{array}{cc} 1 & 1\\ 0 & 1 \end{array} \right ]^{\otimes n}.$$  
The polarization phenomenon states that after applying this linear transformation, every element of the set of conditional entropies $\{H(Y_i|Y_1^{i-1})\}_{i=1}^N$ tends to be either very close to $0$ (fully deterministic) or very close to $1$  (fully random). Moreover, the  fraction of those fully random (informative) variables turns out to be equal to the entropy of the source $h_2(p)$ asymptotically as $N \to \infty$. This allows to build optimal linear source encoders achieving the fundamental information theoretic limit by simply keeping only those rows of $\bfG_N$ corresponding to the fully random variables.

Polar codes have been applied to other problems in communication theory such as as multi-level lattice coding \cite{polar_lattice}, and designing capacity achieving codes over the AWGN channel \cite{abbe_gaussian}, mainly by extending the finite alphabet results. What is less understood is the similar counterpart of the polarization phenomenon for infinite-alphabet sources.
In \cite{discrete_partial_hadamard_isit}, using a new \textit{Entropy Power Inequality} (EPI) for integer-valued random variables \cite{epi_isit}, a novel polarization result was proved for integer-valued sources under the conventional arithmetic over $\bZ$. The result was used to construct deterministic partial Hadamard matrices for almost lossless encoding of integer-valued signals with a vanishing measurement rate of $o(N)$ for large block-lengths $N$. The more general case of real-valued sources, however, was left open in \cite{discrete_partial_hadamard_isit}.
In this paper, we extend this result to real-valued sources and provide a  polarization theory for  infinite-alphabet signals.

\section{R\'enyi information dimension}\label{section:RID}
Let $X$ be a random variable 
 with a probability distribution $p_X$ over $\bR$. The upper and the lower RID of this random variable were defined in \eqref{eq:urid} and \eqref{eq:lrid} respectively. Let $m\in \bN$ and suppose $X \in [0,1]$ almost surely. It is not difficult to see that if $X=0.X_1X_2\dots$ is the $m$-ary expansion of the random variable $X$ with $X_i \in \{0,1,\dots, m-1\}$, then for $q=m^k$, we have $H(\quant{X}_{q})=H(X_1,X_2, \dots, X_k)$, where $H$ denotes the discrete entropy in basis $m$. From \eqref{eq:urid} and \eqref{eq:lrid}, we have
\begin{align}
\lrid{X} \leq {H}_\infty(\{X_i\}_{i=1}^\infty) \leq H^\infty(\{X_i\}_{i=1}^\infty)\leq \urid{X},
\end{align}
where ${H}_\infty=\liminf_{k \to \infty}\frac{H(X_1,X_2, \dots, X_k)}{k}$ denotes the lower entropy rate of the stochastic process $\{X_i\}_{i=1}^\infty$ (with a similar expression for $H^\infty$ by replacing $\liminf$ with $\limsup$).
As a special case, when $X$ is uniformly distributed over $[0,1]$, the random variables $\{X_i\}_{i=1}^\infty$ are i.i.d. each having a uniform distribution over $\{0,1,\dots,m-1\}$. Thus, the upper and lower RID are equal to $\urid{X}=\lrid{X}=1$. Also, $\urid{X}=\lrid{X}=0$ for any discrete random variable $X$ with $H(X)< \infty$. 

By Lebesgue decomposition theorem \cite{halmos2013measure}, any probability distribution $p_X$ over $\bR$ can be written as a convex combination of a continuous part $p_c$, a singular part $p_s$, and a discrete part $p_d$ (with the latter two  being singular with respect to Lebesgue measure) as follows 
\begin{align}\label{leb_dec}
p_X=\alpha_c\, p_c + \alpha_s\, p_s + \alpha_d\, p_d,
\end{align}
where $\alpha_c,\alpha_s,\alpha_d\geq 0$ and $\alpha_c+\alpha_s+\alpha_d=1$. In this paper, we only consider the case $\alpha_s=0$, where $p_X$ is the mixture of a continuous and a discrete distribution. In \cite{renyi}, R\'enyi showed that for such a mixture distribution, the RID is well-defined and is given by the weight of the continuous part $\alpha_c$. In particular, it is $1$ for the continuous and $0$ for the discrete distributions. He also defined the RID of a continuous vector random variable $X_1^n$ of dimension $n$, where he proved that 
\begin{align}
d(X_1^n)=\lim_{q\to \infty} \frac{H(\quant{X_1^n}_q)}{\log_2(q)}=n,
\end{align}
where the quantization is done component-wise, i.e., 
\begin{align}
\quant{X_1^n}_q=(\quant{X_1}_q, \dots, \quant{X_n}_q).
\end{align}

\section{Summary of the Results}
In this section, we briefly explain the results proved in our paper.
Let $Z_1$ and $Z_2$ be two i.i.d. nonsingular random variables with a mixture distribution $p_Z(z) =(1-\delta) p_d(z) + \delta p_c(z)$, where $p_d$ and $p_c$ denote the discrete and the continuous part of $p_Z$. Note that $d(Z_1)=d(Z_2)=\delta$, as explained in Section \ref{section:RID}. 
Let us consider $X_1^2=\bfH_2 Z_1^2$ where $$\bfH_2=\left [\begin{matrix} 1 & 1 \\ 1 & -1 \end{matrix} \right ]$$ denotes the $2\times 2$ Hadamard matrix. A direct calculation shows that $X_1=Z_1+Z_2$ has the following distribution
\begin{align}
p_{X_1}(x) &=p_{Z}(x) \star p_Z(x)= (1-\delta)^2 p_d\star p_d(x) \nonumber\\
& \ + 2\delta (1-\delta) p_d\star p_c(x) + \delta^2 p_c \star p_c(x),\label{conv}
\end{align}
where $\star$ denotes the convolution operator over $\bR$. From \eqref{conv}, it is seen that $p_{X_1}$ is a mixture distribution with a discrete part $(1-\delta)^2 p_d \star p_d$, and has the RID $d(X_1)=1-(1-\delta)^2=2\delta-\delta^2$. 

Now let us consider the conditional distribution of $X_2$ given $X_1$ denoted by $p_{X_2|X_1}(x)$. From standard results in probability theory \cite{chung2001course}, this conditional distribution is a well-defined mixture distribution for almost all realizations of $X_1$. Hence, the conditional RID of $X_2$ given $X_1$ denoted by $d(X_2|X_1=x_1) \in [0,1]$ is well-defined almost surely and is a function of $X_1$. Define the conditional RID of $X_2$ given $X_1$ as $d(X_2|X_1)=\bE_{X_1}[d(X_2|X_1=x_1)]$, provided that $d(X_1|X_2=x_1)$ is a random variable (i.e., a measurable function of $x_1$) with a well-defined expected value. In Section \ref{Gen_RID}, we develop techniques to compute $d(X_2|X_1)$ for a large class of mixture distributions in a closed form, where in particular we prove that such a conditional RID is well-defined. We also extend those techniques to calculate the joint (e.g., $d(X_1,X_2)$), conditional (e.g., $d(X_2|X_1)$), and the mutual RID of mixture vector-valued random variables. As a result, we obtain that $d(X_2|X_1)=\delta^2$ and that 
\begin{align}
d(X_1)+d(X_2|X_1)= (2 \delta -\delta^2) + \delta^2= 2 \delta = 2d(Z_1),
\end{align}
which is analogous to the chain rule for the mutual information \cite{cover2012elements}. In fact, we prove that such a chain rule holds for the RID and satisfies most of the properties of the traditional chain rule for the mutual information \cite{cover2012elements}. In particular,
\begin{align}
d(X_1)+d(X_2|X_1)&=d(X_1,X_2)\stackrel{(i)}{=}d(Z_1,Z_2)\stackrel{(ii)}{=}2d(Z_1),
\end{align}
where $(i)$ follows from the fact that $\bfH_2$ is an invertible matrix, and where $(ii)$ is due to the fact that $Z_1$ and $Z_2$ are i.i.d.

It is, thus, seen that multiplying two i.i.d. random variables $Z_1,Z_2$, with a mixture distribution, by $\bfH_2$ modifies their conditional RIDs according to $(\delta, \delta) \mapsto (2\delta -\delta^2, \delta^2)$. This resembles the  polarization of a BEC channel with a capacity $\delta \in (0,1)$ as in \cite{polar_channel}. In Section \ref{sec:RID_polarization}, we prove that such a polarization indeed occurs for the RID. To be more precise, let $\{Z_i: i \in [N] \}$ be a sequence of i.i.d. nonsingular random variables with an RID $\delta$ and let $X_1^N=\bfH_N Z_1^N$, where $N=2^n$ is a power of two and where $\bfH_N=\left [\begin{matrix} 1 & 1 \\ 1 & -1 \end{matrix} \right ]^{\otimes n}$ denotes the Hadamard matrix of order $N$. We prove that the sequence of conditional RIDs $$\{d(X_i|X_1^{i-1}): i \in [N]\},$$ for increasing values of $N=2^n$,  polarizes according to the polarization pattern of a BEC with a channel capacity $\delta \in (0,1)$. We further investigate the applications of the established polarization result in Section \ref{sec:application}.

\section{Generalization of the RID}\label{Gen_RID}
\subsection{Space of Random Variables and Generalized RID}
Our objective is to extend the definition of RID to vector-valued random variables, which are not necessarily continuous. Let $\clI$ be a collection of independent and nonsingular random variables (with $\alpha_s=0$ as in \eqref{leb_dec}). We define the space $\clL$ of random variables generated by $\clI$ as $\clL=\cup_{n=1}^\infty \clL_n$, where
\begin{align}\label{l_def}
\clL_n=\{X_1^n: \exists\, k, \sfA\in \bR^{n\times k}&, Z_i \in \clI \text{ for } i \in [k],\nonumber\\
&\text{ such that } X_1^n=\sfA Z_1^k\}.
\end{align}
It is seen that $\clL_n$ consists of all $n$-dim random vectors  generated by a linear mixture of \textit{finitely many} elements of $\clI$\footnote{In this paper, we mainly deal with linear transforms of i.i.d. variables, and our main motivation for defining this space is that it remains stable under linear operations. Moreover, using the underlying linear structure, we are able to extend the RID in a natural way to all the variables in this space.}. Note that $\clL$ is stable under vector addition and concatenation, i.e., for arbitrary $W_1^n, X_1^n \in \clL_n$ and $Y_1^m \in \clL_m$, we have that $W_1^n + X_1^n \in \clL_n$, and $[X_1^n; Y_1^m] \in \clL_{n+m}$. Moreover, $\clL$ is stable under an arbitrary linear transformation, i.e., $\psi(\clL_n) \subseteq \clL_m$ for any linear map $\psi: \bR^n \to \bR^m$.

We define the joint and the conditional RID, and the \textit{Mutual R\'enyi Information}  for the random variables in $\lin$  by
\begin{align}
d(X_1^n)&=\lim_{q\to\infty} \frac{H(\quant{X_1^n}_q)}{\log_2(q)}\label{rid_def}\\
d(X_1^n|Y_1^m)&=\lim_{q\to\infty} \frac{H(\quant{X_1^n}_q|Y_1^m)}{\log_2(q)}\label{cond_rid_def}\\
\mri{X_1^n;Y_1^m}&= d(X_1^n)-d(X_1^n|Y_1^m).
\end{align}
We will prove that all the limits above are well-defined. In general, computing the RID for a given multi-variate  distribution is quite challenging since the distribution might contain a probability mass over complicated subsets or sub-manifolds of lower dimensions. Moreover, the limit might not even exist in some cases.  Fortunately, using the linear structure in $\clL$, we are able to obtain a simple formula for computing the RID via the rank characterization. {A similar rank characterization was used in the context of finite fields for coding over the BEC and the BSC (\textit{Binary Symmetric Channel}) in \cite{abbe2015high}}. We first need some notation and definitions.

\begin{definition}\label{res_def}
Let $\sfA$ and $\sfB$ be two arbitrary matrices of dimension $m_a\times n$ and $m_b\times n$, and let $\clC \subseteq [n]$. The residual of matrix $\sfA$ given $\sfB$ over the column set $\clC$ is defined by \begin{align}\label{eq:res_def}
\res[\sfA|\sfB;\clC]&=\rank([\sfA;\sfB]_\clC)-\rank(\sfB_\clC).
\end{align} 
\end{definition}
It is seen that $\res[\sfA|\sfB;\clC]$ is  the amount of increase in the rank of $\sfB_\clC$ by adding the rows of $\sfA_\clC$. In particular, if the rows of $\sfA_\clC$ are in the row-span of $\sfB_\clC$, then $\res[\sfA|\sfB;\clC]$ is $0$.
\begin{example}
Let $\sfA=[1, 1]$ and $\sfB=[1, 0]$. Then, 
\begin{align}
\res[\sfA|\sfB; {\emptyset}]=0&, \ \res[\sfA|\sfB; {\{1\}}]=0,\\
\res[\sfA|\sfB; {\{2\}}]=1&, \ \res[\sfA|\sfB; {\{1,2\}}]=1.
\end{align}
\end{example}
Many properties of $\res$ simply follow from the algebraic properties of the rank. In this paper, we need additionally the following properties of $\res$ summarized in Proposition \ref{prop:res_prop}. 
\begin{proposition}\label{prop:res_prop}
Let $\sfA$, $\check{\sfA}$ be $m_a\times n$ matrices. 
The operator $\res$ satisfies the following properties:
\begin{itemize}
\setlength{\itemindent}{-2mm}
\item ({chain rule}) Let \{$\clR_i\}_{i=1}^p$ with $\clR_i \subseteq [m_a]$ be an arbitrary partition of the rows of $\sfA$, and let $\clC\subseteq [n]$. Then,
\begin{align}
\sum_{i=1}^p \res\Big[\sfA[\clR_i]\big|\sfA[\cup_{\ell=1}^{i-1} \clR_\ell]; \clC\Big ]=\rank(\sfA_\clC).
\end{align}

\item ({rank-$1$ innovation}) Let $\sfa, \check{\sfa} \in \bR^{1\times n}$. Suppose $\clC, \check{\clC} \subseteq [n]$ are arbitrary subsets of the columns of $\sfA$ and $\check{\sfA}$. Then,
\begin{align}\label{rank1_innov}
\res\Big[[\sfa, \check{\sfa}]\big |[\sfA, \check{\sfA}];&\,  {\clC\sqcup \check{\clC}}\Big]\geq\res\big [\sfa|\sfA; {\clC}\big] + \res\big[\check{\sfa}|\check{\sfA}; {\check{\clC}}\,\big] \nonumber\\
&- \res\big[\sfa|\sfA; {\clC}\big]\res\big[\check{\sfa}|\check{\sfA}; \, {\check{\clC}}\,\big],
\end{align}
where $\sqcup$ denotes the disjoint union. The equality holds in \eqref{rank1_innov} if $\sfA$ and $\check{\sfA}$ have non-overlapping set of nonzero rows, i.e., $\sfA$ has zero rows in the row-set corresponding to the nonzero rows of $\check{\sfA}$ and vice versa.  \hfill {\LARGE $\square$}
\end{itemize}
\end{proposition}
\begin{proof}
Proof in Appendix \ref{prop:res_prop_app}.
\end{proof}

\subsection{Properties of the RID over $\clL$}\label{sec:RID_prop}
We first need some notation to  simplify the statement of the results in this section. Let $X_1^n \in \clL_n$ and $Y_1^m \in \clL_m$ be random vectors in $\clL$. From the definition in \eqref{l_def}, there are matrices $\sfA$ and $\sfB$ of dimension\footnote{By adding zero columns whenever needed, without loss of generality, we can always assume that $\sfA$ and $\sfB$ have the same number of columns.} $n \times k$ and $m\times k$ for some finite $k$ and independent nonsingular random variables $Z_1^k \in \clI$, such that
 $X_1^n=\sfA Z_1^k$ and $Y_1^m=\sfB Z_1^k$. Since each $Z_i$ has a mixture distribution, it can be represented as $Z_i=\Theta_i \contin_i + (1-\Thet_i) \disc_i$, where $\contin_i \sim p_{c_i}$ and $\disc_i \sim p_{d_i}$ denote the continuous and the discrete part of $Z_i$ and their corresponding distributions over $\bR$, and where $\Thet_i \in \{0,1\}$ is a binary random variable independent of $\contin_i$ and $\disc_i$ with $\bP[\Theta_i=1]=d(Z_i)$. We define the support set of the random vector $Z_1^k$ by 
\begin{align}\label{eq:supp_set}
\clC=\{i\in[k]: \Theta_i=1\}.
\end{align}
It is seen that $\clC$ is a random subset of $[k]$. Moreover, $\bP[i \in \clC]=\bP[\Theta_i=1]=d(Z_i)$, thus, $\clC$ has the average cardinality $\bE[|\clC|]=\sum_{i=1}^k d(Z_i)$. We have the following result. 
 \begin{theorem}\label{RID_maintheorem}
Let $(X_1^n,Y_1^m)$ and $\clC$ be as before. Then,  
\begin{itemize}
\item $d(X_1^n)=\bE [ \rank(\sfA_\clC)]$,
\item $d(X_1^n|Y_1^m)=\bE \big[ \res[\sfA|\sfB;\clC]\big]$,
\end{itemize}
with the expectation taken over the random support set $\clC$.  \hfill {\LARGE $\square$}
\end{theorem}
\begin{proof}
Proof in Appendix \ref{RID_maintheorem_app}.
\end{proof}

\begin{remark}\label{disc_comp_remark}
Note that if one of the variables in $Z_1^k$, say $Z_1$, is discrete, then $\bP[\Thet_1=1]=0$, which implies that $1\notin \clC$. Hence, the first column of the matrices $\sfA$ and $\sfB$ will never be selected. From Theorem \ref{RID_maintheorem}, this implies that we can drop the fully discrete constituents of $X_1^n$ and $Y_1^m$ (e.g., $Z_1$ here) without changing their individual or joint RIDs. 
\hfill $\Diamond$
\end{remark}

Using Theorem \ref{RID_maintheorem} and the properties of the $\res$ operator in Proposition \ref{prop:res_prop}, we obtain the following properties of the RID. 
\begin{theorem}\label{RID_extensions}
Let $(X_1^n,Y_1^m)$ be a random vector in $\lin$ as in Theorem \ref{RID_maintheorem}. Then, we have the following properties:
\begin{itemize}
\setlength{\itemindent}{-2mm}
\item ({positivity}) $d(X_1^n)\geq 0$, with the equality  if and only if every $X_i$, $i \in [n]$, is discrete.
\item ({invariance}) $d(X_1^n)=d(\sfL X_1^n)$ for any invertible $n\times n$ matrix $\sfL$.
\item ({chain rule}) $d(X_1^n,Y_1^m)=d(X_1^n)+d(Y_1^m|X_1^n)$.
\item ({symmetry}) $\mri{X_1^n;Y_1^m}=\mri{Y_1^m;X_1^n}$.
\item ({positivity}) $\mri{X_1^n;Y_1^m}\geq 0$. \hfill {\LARGE $\square$}
\end{itemize}
\end{theorem}
\begin{proof}
Proof in Appendix \ref{RID_extensions_app}.
\end{proof}

\begin{example}
Let $Z_1^3$ be i.i.d. with $d(Z_i)=0.6$ for $i=1,2,3$. Let $X=Z_1+Z_2$ and $Y=Z_2+Z_3$. This can be written in the following form 
\begin{align}
\left [ \begin{array}{c} X \\ Y \end{array} \right ]= \left [ \begin{array}{ccc} 1 & 1 & 0\\ 0 & 1 & 1 \end{array} \right ]\left [ \begin{array}{c} Z_1\\ Z_2 \\ Z_3 \end{array} \right ]=\sfA Z_1^3.
\end{align}
For computing $d(X)$, let $\sfa=[1,1,0]$ denote the first row of $\sfA$. It is seen that for any $\clC \subseteq [3]$, the rank of $\sfa_\clC$ is equal to $1$ except when $\clC=\emptyset$ or $\clC=\{3\}$. Thus,
\begin{align}
d(X)&=1-\bP[\clC=\emptyset] - \bP[\clC=\{3\}]\\
&= 1- 0.4^3 - (0.4)^2 (0.6)=0.84.
\end{align}
From symmetry, we also have $d(Y)=0.84$. To compute $d(X,Y)$, we can see that for any $\clC \subseteq [3]$, the rank of $\sfA_\clC$ is equal to $|\clC|$ except when $\clC=[3]$, where all the columns are selected. Hence, $\rank(\sfA_\clC)=|\clC| - \ind_{\{\clC=[3]\}}$, which gives 
\begin{align}
d(X,Y)&=\bE[|\clC|]- \bP[\clC=[3]]\\
&= 3\times 0.6 - 0.6^3= 1.584.
\end{align}
Using the chain rule for the RID, we obtain $d(X|Y)=d(Y|X)= 1.584- 0.84=0.744$, where it is seen that $d(X|Y) < d(X)$. We can also directly compute $d(X|Y)$ using 
\begin{align}
d(X|Y)=\bE \{ \res[\sfa|\sfb;\clC]\},
\end{align}
 where $\sfb$ denotes the second row of $\sfA$. We can simply check that $\res[\sfa|\sfb;\clC]=1$ except when $\clC\in \Big \{\emptyset, \{2\}, \{3\} \Big \}$. Hence,
 \begin{align}
 d(X|Y)&= 1- \bP\Big[\clC\in \Big \{\emptyset, \{2\}, \{3\} \Big \}\Big]\\
 &= 1-( 0.4^3 + 2\times 0.6 (0.4)^2)\\
 &= 1-0.256=0.744.
 \end{align}
The Mutual R\'enyi Information between $X$ and $Y$ is given by 
\begin{align}
\mri{X;Y}=d(X)-d(X|Y)= 0.84-0.744=0.096.
\end{align}
\end{example}

\section{Polarization of the RID}\label{sec:RID_polarization}
\subsection{Basic Definitions and Results}\label{rid_basic}
Before stating the polarization result for the RID, we first define the \textit{erasure process}.

\begin{definition}\label{eras_proc_def}
Let $\alpha \in [0,1]$. An ``erasure process'' with an initial value $\alpha$ is defined as follows:
\begin{enumerate}
\item $e^\emptyset=\alpha$. $e^+=2\alpha-\alpha^2$ and $e^-=\alpha^2$.
\item Let $e_n:=e^{b_1b_2 \dots b_n}$ for some $\{+,-\}$-valued sequence $b_1^n$. Define 
\begin{align*}
e_n^+&:=e^{b_1b_2\dots b_n+}=2e_n - e_n^2,\\
e_n^-&:=e^{b_1b_2\dots b_n-}=e_n^2. 
\end{align*}                       
\end{enumerate}
\end{definition}
Using the $\{+,-\}$ labeling, we can construct a binary tree where each leaf of the tree is labeled with a specific $\{+,-\}$-valued sequence and is assigned the erasure value corresponding to the same $\{+,-\}$-valued sequence.

Let $\{B_n\}_{n=1}^\infty$ be a sequence of i.i.d. uniformly distributed $\{+,-\}$-valued random variables. By replacing $B_1^n$ for $\{+,-\}$-labeling $b_1^n$ in the definition of the erasure process, we obtain a stochastic process $E_n=e^{B_1B_2 \dots B_n}$. Let $\clF_n$ be the $\sigma$-field generated by $B_1^n$. The BEC polarization  can be summarized as follows \cite{polar_channel, rate_polar}:
\begin{enumerate}
\item $(E_n,\clF_n, \bP)$ is a positive martingale bounded in $[0,1]$.
\item $E_n$ converges to $E_\infty \in \{0,1\}$ with $\bP(E_\infty=1)=\alpha$.
\item For any $\beta \in (0, \frac{1}{2})$,
\begin{align}\label{alp_erasure}
\liminf _{n \to \infty} \bP(E_n \geq 1- 2^{-2^{n\beta}})=\alpha,\\
\liminf_{n \to \infty} \bP(E_n \leq 2^{-2^{n\beta}})=1-\alpha.
\end{align}
\end{enumerate}

\subsection{RID Polarization}\label{sec:rid_polar}
Let $N=2^n$ be a power of $2$ and let $Z_1^N$ be a sequence of i.i.d. nonsingular random variables with an RID $d(Z_i)=\delta\in (0,1)$. Let ${\bH}_N$ be a Hadamard matrix of order $N$ with the following recursive relation between ${\bH}_N$ and ${\bH}_{2N}$
\begin{align}\label{had_recursive}
\begin{array}{ccc}
{\bH}_N=\left [
\begin{array}{c}{\sfh}_1  \\
\vdots \\
{\sfh}_N  \\
\end{array} \right ]
& \to &
{\bH}_{2N}=\left [ 
\begin{matrix} {\sfh}_1 &,&{\sfh}_1\\
\,{\sfh}_1 &,& -{\sfh}_1\\
\vdots &,&\vdots\\
{\sfh}_i &,& {\sfh}_i\\
{\sfh}_i &,& -{\sfh}_i\\
\vdots &,&\vdots\\
\end{matrix} \right ]
 \end{array},
\end{align}
where ${\sfh}_i$, $i\in[N]$, denotes the $i$-th row of ${\bH}_N$. 
This corresponds to a standard Hadamard matrix with shuffled rows. This construction simplifies the proofs, but all the result are still valid for the standard Hadamard matrix $\bfH_N$ 
without any shuffling. 

Let ${\bH}_N$ be as in \eqref{had_recursive} and let $X_1^N={\bH}_N Z_1^N$ be the vector of variables obtained by the Hadamard transform of $Z_1^N$. Let us define 
\begin{align}\label{I_def}
I_n: [N] \to [0,1], \ I_n(i)=d(X_i|X_1^{i-1}), i\in [N].
\end{align}
Assume that $b_1^n$ is the binary expansion of $i-1$. By replacing $0$ by $+$ and $1$ by $-$, we can equivalently label $I_n(i)$ by a sequence of $\{+,-\}$ of length $n$, i.e., $I_n(i)=I^{b_1b_2\dots b_n}$. Similar to the erasure process, we can convert $I_n$ to a stochastic process $I_n=I^{B_1B_2\dots B_n}$ by using i.i.d. uniform $\{+,-\}$-valued random variables $B_1^n$. We can now prove the following theorem.

\begin{theorem}[RID Polarization]\label{single_RID_polarization}
$(I_n,\clF_n,\bP)$ is an erasure stochastic process with initial value $\delta$ polarizing to $\{0,1\}$. \hfill {\LARGE $\square$}
\end{theorem}
\begin{proof}
For $n=0$, we have a Hadamard matrix of order $N=2^0=1$ which is simply a number, thus, $X_1=Z_1$ and we have $I_0(1)=d(Z_1)=\delta$. Consider an arbitrary $n$, let $N=2^n$ and let $I_n$ be defined as in \eqref{I_def}. We need to prove that $I_n$ satisfies the following recursion for $i \in [2^n]$
\begin{align}\label{i_evolution}
I_n(i)^+&=I_{n+1}(2i-1)= 2 I_n(i)-I_n(i)^2\\
I_n(i)^-&=I_{n+1}(2i)=I_n(i)^2.
\end{align}
As $Z_1^N$ are i.i.d. nonsingular random variables, it results that $X_1^N={\bH}_N Z_1^N$ belongs to the space $\clL$ generated by $Z_1^N$. Hence, 
using the rank characterization for the RID over $\lin$ in Theorem \ref{RID_maintheorem}, we have  
\begin{align}\label{I_rank_def1}
I_n(i)=d({X}_i|{X}_1^{i-1})=\bE_\clC \big [\res[\sfh_i|{\bH}_N[\interv{1}{i-1}];\clC]\big ],
\end{align}
where ${\bH}_N[\interv{1}{i-1}]$ denotes the $(i-1)\times N$ matrix consisting of the first $i-1$ rows of ${\bH}_N$ and where $\clC$ denotes the random support of continuous parts of $Z_1^N$ as defined in \eqref{eq:supp_set}. Recall that $i\in \clC$ if and only if the random variable $Z_i$ is sampled according to the continuous part of its distribution.

At stage $n+1$, we have the term $I_n(i)^+$ which corresponds to the row $2i-1$ of ${\bH}_{2N}$ as follows
\begin{align}\label{rec_rank_1}
{\bH}_{2N}[\interv{1}{2i-1}]=\left [ 
\begin{matrix} {\sfh}_1 &,&{\sfh}_1\\
\,{\sfh}_1 &,& -{\sfh}_1\\
\vdots &,&\vdots\\
\sfh_{i-1} &,& \sfh_{i-1}\\
\sfh_{i-1} &,& - \sfh_{i-1}\\
 {\sfh_i} &,&  {\sfh_i}
\end{matrix} \right ],
\end{align}
where $I_n(i)^+$ is defined similar to \eqref{I_rank_def1} by
\begin{align}\label{I_rank_def2}
I_n(i)^+=\bE_{\clC, \check{\clC}} \Big [\res\Big [[\sfh_i, \sfh_i] \big|{\bH}_{2N}[\interv{1}{2(i-1)}];\clC\sqcup \check{\clC}\Big ]\Big ],
\end{align}
where $\clC$ and $\check{\clC}$ denote the support set of $Z_1^N$ and $\check{Z}_1^N:= Z_{N+1}^{2N}$. As $\check{Z}_1^N$ is an independent copy of $Z_1^N$, the support sets $\clC$ and $\check{\clC}$ are independent and identically distributed. Applying a simple row operation to ${\bH}_{2N}[\interv{1}{2(i-1)}]$, which preserves the rank, we have that 
{\small 
\begin{align*}
\res\Big[[\sfh_i, \sfh_i] \big|{\bH}_{2N}[\interv{1}{2(i-1)}];\clC\sqcup\check{\clC}\Big]=\res\Big[[\sfh_i, \sfh_i] \big|[\sfL, \sfR];\clC\sqcup\check{\clC}\Big],
\end{align*} 
}
where $\sfL$ and $\sfR$ are  given by 
\begin{align}\label{LR_mat}
\sfL=\left [ 
\begin{matrix} {\sfh}_1\\
{\bf 0}\\
\vdots \\
\sfh_{i-1} \\
{\bf 0}
\end{matrix} \right ],
\sfR=\left [ 
\begin{matrix} \ \ {\bf 0}\\
-{\sfh}_1\\
\vdots \\
\ \ {\bf 0}
 \\
-\sfh_{i-1}
\end{matrix} \right ].
\end{align}

\begin{figure*}[t]
\centering
\includegraphics{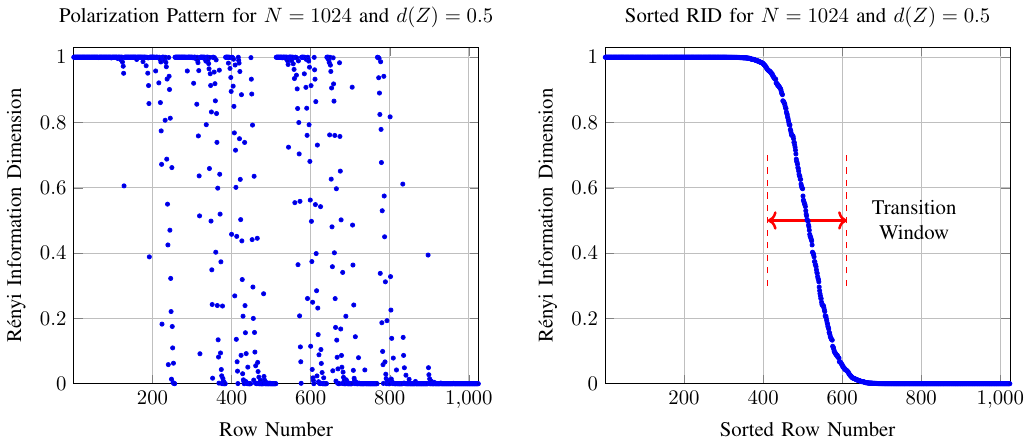}
\caption{Polarization pattern of an i.i.d. source $\{Z_i\}_{i=1}^\infty$ with $d(Z_1)=0.5$ after being transformed by the Hadamard matrix of order $N=1024$. It is seen that most of the rows are polarized to their corresponding RIDs, and the fraction of those rows polarized to $1$ converges to $d(Z_1)=0.5$.}
\label{polar1024}
\end{figure*}

Since $\sfL$ and $\sfR$ have non-overlapping rows, using the rank-$1$ innovation property in Proposition \ref{prop:res_prop}, we obtain that
\begin{align}\label{I_rank_res}
\res\Big[[\sfh_i, \sfh_i] \big|[\sfL, \sfR];\clC\sqcup\check{\clC}\Big]&=\res\Big[\sfh_i\big|\sfL;\clC\Big]+ \res\Big[\sfh_i\big|\sfR;\check{\clC}\Big] \nonumber\\
&- \res\Big[\sfh_i\big|\sfL;\clC\Big] \res\Big[\sfh_i\big|\sfR;\check{\clC}\Big].
\end{align}
From \eqref{LR_mat}, it is also seen that 
\begin{align}
\res[h_i|\sfL;\clC]&=\res\Big[h_i\big|\bH_{N}[\interv{1}{i-1}];\clC\Big],\label{I_rank_def1_1}\\
\res[h_i|\sfR;\check{\clC}]&=\res\Big[h_i\big|\bH_{N}[\interv{1}{i-1}];\check{\clC}\Big]\label{I_rank_def1_2}.
\end{align}
As $\clC$ and $\check{\clC}$ are i.i.d., taking the expectation from \eqref{I_rank_res}, and using \eqref{I_rank_def1_1}, \eqref{I_rank_def1_2}, and also \eqref{I_rank_def1}, we have
\begin{align}\label{I_plus}
I_n(i)^+&= I_n(i) + I_n(i) - I_n(i) \cdot I_n(i)\nonumber\\
&=2I_n(i) - I_n(i)^2,
\end{align}
which proves the first identity in \eqref{i_evolution}. To prove the second identity in \eqref{i_evolution}, let $\check{Z}_1^N$ and $Z_1^N$ be independent copies as defined before and let $W_1^N={\bH}_N Z_1^N$ and $\check{W}_1^N={\bH}_N \check{Z}_1^N$. We obtain that 
\begin{align}
X_{2i-1}=W_i+\check{W}_i,\ 
X_{2i}=W_i -\check{W}_i.
\end{align}
Moreover, by definition, we have
\begin{align*}
I_n(i)^+&=d(X_{2i-1}|X_1^{2i-2})=d(W_i+\check{W}_i | W_1^{i-1}, \check{W}_1^{i-1})\\
I_n(i)^-&=d(X_{2i}|X_1^{2i-1})\\
&=d(W_i-\check{W}_i | W_1^{i-1}, \check{W}_1^{i-1}, W_i+\check{W}_i).
\end{align*}
Applying the chain rule for RID from Theorem \ref{RID_extensions} and using the independence of $W_1^N$ and $\check{W}_1^N$, we obtain
\begin{align}\label{ch_rule_single}
I_n(i)^+ + I_n(i)^-&= d(W_i+\check{W}_i, W_i-\check{W}_i|W_1^{i-1}, \check{W}_1^{i-1})\nonumber\\
&=d(W_i, \check{W}_i|W_1^{i-1}, \check{W}_1^{i-1})\nonumber\\
&=2 d(W_i|W_1^{i-1})=2I_n(i).
\end{align}
From \eqref{I_plus}, this implies $I_n(i)^-=I_n(i)^2$, which proves the second identity in \eqref{i_evolution}. From Definition \ref{eras_proc_def}, this indicates that $I_n$ is an  erasure process with the initial value $d(Z_1)=\delta$. 
\end{proof}

Fig.\,\ref{polar1024} illustrates the polarization pattern for an i.i.d.  source $\{Z_i\}_{i=1}^\infty$ with $d(Z_1)=0.5$ after being transformed by the Hadamard matrix of order $N=2^{10}=1024$. It is seen that even for $N=1024$ more than $80\%$ of the rows are polarized to their corresponding RIDs.

\subsection{RID-Preserving Matrices}\label{rid_pre_single}
Let $\{Z_i\}_{i=1}^\infty$ be an i.i.d. source with $d(Z_1)\in (0,1)$. 
Let $\{\sfA^{(k)}\}_{k=1}^\infty$ be a sequence of matrices of order $m(k)\times k$. We say that  $\{\sfA^{(k)}\}_{k=1}^\infty$ is an RID-preserving family for $\{Z_i\}_{i=1}^\infty$ if and only if \begin{align}\label{rid_preserve}
d(\sfA^{(k)} Z_1^k) \geq d(Z_1^k)-o(k),
\end{align}
 where $o(k)$ denotes a vanishing term compared with $k$ as  $k$ tends to infinity. We define the  asymptotic measurement rate of the family $\{\sfA^{(k)}\}_{k=1}^\infty$ by $\rho:=\limsup_{k\to \infty} \frac{m(k)}{k}$. From \eqref{rid_preserve}, it is seen that taking measurements with this family of matrices asymptotically preserves the whole RID of the source.
\begin{proposition}\label{rid_pres_prop}
Let $\{Z_i\}_{i=1}^\infty$ be an i.i.d. source with a given $d(Z_1) \in (0,1)$ and let $\{\sfA^{(k)}\}_{k=1}^\infty$ be an RID-preserving family for the source. Then, $\rho \geq d(Z_1)$. \hfill {\LARGE $\square$}
\end{proposition} 
\begin{proof}
First note that for every $k$, the random vector $\sfA^{(k)} Z_1^k$ belongs the space $\clL$ generated by the i.i.d. nonsingular variables $\{Z_i\}_{i=1}^\infty$, thus, it has a well-defined RID. In particular, if $\clC$ is the support set of the i.i.d. nonsingular variables $Z_1^k$, as defined in \eqref{eq:supp_set},  from the rank property of the RID proved in Theorem \ref{RID_maintheorem}, we have 
\begin{align}
m(k) &\geq \bE[\rank(\sfA^{(k)}_\clC)] =d(\sfA^{(k)} Z_1^k)\\
&\geq d(Z_1^k) - o(k)=k d(Z_1) -o(k).
\end{align}
Dividing both sides by $k$, and taking the limit as $k$ tends to infinity, we obtain the desired result $\rho \geq d(Z_1)$.
\end{proof}
Proposition \eqref{rid_pres_prop} implies that to be RID-preserving, any family of matrices needs to have a measurement rate at least as large as the RID of the source. 
To show that this measurement rate is indeed sufficient, we build an RID-preserving family $\{\sfA^{(k)}\}$, labeled with $k \in \{\interv{2^n}{\ n \in \bN}\}$, where $\sfA^{(k)}$ is a submatrix of the Hadamard matrix $\bH_k$, obtained by selecting specific rows of $\bH_k$. We then prove that the measurement rate of the constructed family is $d(Z_1)$.

Let $N=2^n$ be a power of two, let $\bH_N$ be the shuffled Hadamard matrix of order $N$  defined in \eqref{had_recursive}, and let $X_1^N=\bH_N Z_1^N$. Set $\beta \in (0,\frac{1}{2})$, and $\epsilon^{(N)} =2^{-N^\beta}$, and let $\sfH^{(N)}$ be a submatrix of $\bH_N$ obtained by selecting those rows of $\bH_N$ belonging to the index set 
\begin{align}\label{sampling_pattern_1}
\clS^{(N)}=\{i\in[N]: d(X_i|X_1^{i-1})\geq  \epsilon^{(N)}\}.
\end{align}
Let $m(N)=|\clS^{(N)}|$ be the number of rows of $\sfH^{(N)}$ and let $\{\sfH^{(N)}\}$ be the resulting family of matrices indexed with $N$, where $N$ is a power of $2$. 
\begin{proposition}\label{single_H_rid_pres}
The sequence of matrices $\{\sfH^{(N)}\}$ is RID-preserving for the i.i.d. source $\{Z_i\}_{i=1}^\infty$ and has a measurement rate $\rho=d(Z_1)$. \hfill {\LARGE $\square$}
\end{proposition}
\begin{proof}
Let $X_1^N=\bH_N Z_1^N$. Note that from Theorem \ref{single_RID_polarization}, the sequence of conditional RIDs $\{d(X_i|X_1^{i-1})\}_{i=1}^N$ is an erasure process with  an initial value $d(Z_1)$. Thus, applying the polarization rate result in \eqref{alp_erasure}, we obtain that for a $\beta \in (0,\frac{1}{2})$ and $\epsilon^{(N)}=2^{-N^\beta}$, the fraction of those conditional RIDs with a value larger than $\epsilon^{(N)}$, i.e., those belonging to $\clS^{(N)}$,  must converge to $d(Z_1)$. This implies that 
$\rho=\limsup_{N \to \infty} \frac{m(N)}{N}=d(Z_1)$. 
To prove the RID-preserving property, let $\sfr(\ell) \in \clS^{(N)}$ be the index of the $\ell$-th row of ${\sfH^{(N)}}$ among the rows of $\bH_N$. Then, we have
\begin{align}\label{rid_pres_confirm}
d(\sfH^{(N)} Z_1^N)&=\sum_{\ell=1}^{m(N)} d(X_{\sfr(\ell)} | X_{\sfr(1)}, X_{\sfr(2)}, \dots, X_{\sfr(\ell-1)})\nonumber\\
&\stackrel{(a)}{\geq} \sum_{\ell=1}^{m(N)} d(X_{\sfr(\ell)} | X_1^{\sfr(\ell)-1})\nonumber\\
&\stackrel{(b)}{=} d(X_1^N) - \sum_{s \notin \clS^{(N)}} d(X_s|X_1^{s-1})\nonumber\\
&\geq d(Z_1^N) - (N- |\clS^{(N)}|)\epsilon^{(N)}\nonumber\\
& \geq d(Z_1^N) - o(N),
\end{align}
where in $(a)$ we used the positivity of the Mutual R\'enyi Information proved in Theorem \ref{RID_extensions} and the fact that conditioning reduces the RID, and where in $(b)$ we used the chain rule for the RID given by $d(X_1^N)=\sum_{s=1}^N d(X_s|X_1^{s-1})$. The Eq.~\eqref{rid_pres_confirm} confirms the RID-preserving property of $\{\sfH^{(N)}\}$. This completes the proof.
\end{proof}

\section{Multi-terminal (Distributed) Polarization}
\subsection{Multi-terminal RID Polarization}\label{multi_rid_polar}
The RID polarization proved in Theorem \ref{single_RID_polarization} can be extended to the multi-terminal signals. Let $\{(U_i,V_i)\}_{i=1}^\infty$ be a sequence of  i.i.d. 2-dim vectors in $\clL$. Since $[U_1, V_1] \in \clL$, there are $\sfa, \sfb \in \bR^{1\times k}$ and i.i.d. nonsingular variables $Z_1^k$ such that $[U_1;V_1]=[\sfa; \sfb] Z_1^k$.
Let $N=2^n$ be power of $2$ and let $X_1^N= \bH_N U_1^N$ and $Y_1^N=\bH_N V_1^N$, where $\bH_N$ is as in \eqref{had_recursive}. In the \textit{single-terminal} case in Section \ref{sec:rid_polar}, we used the chain rule for the variables $X_1^N=\bH_N Z_1^N$ to expand $d(X_1^N)$ in terms of the conditional RIDs $\{d(X_i|X_1^{i-1})\}_{i=1}^N$, thus, obtaining an erasure process with initial value $d(Z_1)$ polarizing to $\{0,1\}$. In the multi-terminal case, however, we obtain different erasure processes by applying the chain rule to $d(X_1^N,Y_1^N)$ with different expansion orders. For example, if we expand first in terms of $X_1^N$ and then in terms of $Y_1^N$, we obtain the following two sequences for $i\in[N]$:
\begin{align}
I_n(i)=d(X_i|X_1^{i-1}), J_n(i)=d(Y_i|Y_1^{i-1}, X_1^N).
\end{align}
To show that $I_n$ and $J_n$ are indeed erasure processes, similar to Section \ref{sec:rid_polar}, we label different components of $I_n$ and $J_n$ with $\{+,-\}$-valued sequences. We remove the details for brevity. We obtain the following result.
\begin{theorem}\label{multi_RID_polarization}
$(I_n, \clF_n , \bP)$ and $(J_n, \clF_n , \bP)$ are erasure processes with initial value $d(U_1)$ and $d(V_1|U_1)$ respectively, polarizing to $\{0,1\}$. \hfill {\LARGE $\square$}

\end{theorem}
\begin{proof}
Proof in Appendix \ref{multi_RID_polarization_app}.
\end{proof}
\noindent By changing the order of expansion of $d(X_1^N, Y_1^N)$, i.e., first expanding with respect to $Y_1^N$ and then with respect to $X_1^N$, we obtain another $2$-dim erasure process $(I_n, J_n)$ with the initial value $(d(U_1|V_1), d(V_1))$, rather than $(d(U_1), d(V_1|U_1))$. In fact, by applying the monotone chain rule expansion introduced in \cite{bilkent2012polar}, we can expand $d(X_1^N, Y_1^N)$ jointly (and simultaneously) in terms of $X$s and $Y$s, thus, we can construct different $2$-dim polarizing erasure processes $(I_n,J_n)$ that converge almost surely to $(I_\infty, J_\infty)\in \{0,1\}^2$. 
Also, the closure of the region of all possible $(\bar{I}, \bar{J}):=\bE[(I_\infty, J_\infty)]$ for polarizing processes $(I_n,J_n)$ contains the dominant face of the $2$-dim region given by
\begin{align}\label{rate_region_2d}
\clR_2= \{\boldsymbol{\rho} \in \bR_+^2: \ &\rho_1\geq d(U_1|V_1), \rho_2\geq d(V_1|U_1), \nonumber\\
&\rho_1+\rho_2\geq d(U_1,V_1)\},
\end{align}
which is a line connecting two points $(d(U_1),d(V_1|U_1))$ and $(d(U_1|V_1), d(V_1))$ in $\bR_+^2$. This resembles  the Slepian-Wolf region for the distributed source coding \cite{slepian}. The results can  be extended to an i.i.d. sequence of $d$-dim signal $\bfU=(U_1,U_2, \dots, U_d)$, for some $d\geq 3$. By applying the chain rule in different orders and using the results in \cite{bilkent2012polar}, it is possible to build a $d$-dim erasure process $(I^{(1)}_n, I^{(2)}_n, \dots, I^{(d)}_n)$ that converges almost surely to $(I^{(1)}_\infty, I^{(2)}_\infty, \dots, I^{(d)}_\infty) \in \{0,1\}^d$. Moreover, the closure of the region of all $d$-dim averages $\bE[(I^{(1)}_\infty, I^{(2)}_\infty, \dots, I^{(d)}_\infty)]$ corresponds to the dominant face of the region  
\begin{align}\label{rate_region_dd}
\clR_d=\{\boldsymbol{\rho} \in \bR_+^d: \sum_{i\in \clT} \rho_i \geq d(\bfU_\clT|\bfU_{\clT^c}), \ \forall \clT \subseteq [d]\},
\end{align}
where $\bfU_\clT$ denotes the subvector of $\bfU$ obtained by selecting the components in $\clT \subseteq [d]$. 

\subsection{RID-Preserving Matrices for Multi-terminal Signals}\label{rid_preserv_multi_terminal}
The RID-preserving property in Section \ref{rid_pre_single} can also be extended in a natural way to multi-terminal signals. For simplicity, we focus on the $2$-dim case. The results can be extended to dimensions larger than $2$. 

Let $\{(U_i,V_i)\}_{i=1}^\infty$ be a sequence of i.i.d. $2$-dim signals belonging to $\clL$ and let $\{\sfA^{(k)},\sfB^{(k)}\}_{k=1}^\infty$ be a sequence of matrices of order $m_a(k)\times k$ and $m_b(k)\times k$. We call $\{\sfA^{(k)},\sfB^{(k)} \}_{k=1}^\infty$ an RID-preserving sequence for $\{(U_i, V_i)\}_{i=1}^\infty$ if and only if 
\begin{align}\label{rid_preserve_multi}
d(\sfA^{(k)} U_1^k, \sfB^{(k)} V_1^k ) \geq d(U_1^k, V_1^k)-o(k),
\end{align}
where $o(k)$ denotes a term vanishing in the dimension $k$. We define $\rho_a:=\limsup_{k\to \infty} \frac{m_a(k)}{k}$, and $\rho_b:=\limsup_{k\to \infty} \frac{m_b(k)}{k}$ the asymptotic measurement rate of the family. We obtain the following result.
\begin{theorem}\label{rid_pres_prop_multi}
Let $\{(U_i,V_i)\}_{i=1}^\infty$ be an i.i.d. source with the joint RID $d(U_1,V_1)$ and conditional RIDs $d(U_1|V_1)$ and $d(V_1|U_1)$.  Let $\{\sfA^{(k)}, \sfB^{(k)}\}_{k=1}^\infty$ be a family of RID-preserving matrices for the source. Then,
\begin{align}\label{rate_region_2d}
\rho_a \geq d(U_1|V_1), \rho_b\geq d(V_1|U_1), \rho_a+\rho_b \geq d(U_1,V_1).
\end{align}
\end{theorem} 
\begin{proof}
First note that from the rank property for the RID proved in Theorem \ref{RID_maintheorem} and the RID-preserving property in \eqref{rid_preserve_multi}, it  results that 
\begin{align*}
m_a(k)+m_b(k) \geq d(\sfA^{(k)} U_1^k, \sfB^{(k)} V_1^k ) \geq d(U_1^k, V_1^k)-o(k),
\end{align*}
simply because the rank of $[\sfA^{(k)};\sfB^{(k)}]$ is less than its number of rows $m_a(k)+m_b(k)$. This implies that $\rho_a+\rho_b \geq d(U_1,V_1)$.
To prove the other two inequalities, note that the RID-preserving property in \eqref{rid_preserve_multi} can be  written as 
\begin{align}
d(U_1^k, V_1^k|\sfA^{(k)}U_1^k, \sfB^{(k)}V_1^k)\leq o(k).
\end{align}
Applying the chain rule, we have
\begin{align} 
d(U_1^k|\sfA^{(k)}U_1^k, \sfB^{(k)}V_1^k)+d(V_1^k| \sfB^{(k)}V_1^k, U_1^k)\leq o(k).
\end{align}
Using the positivity of the RID, this immediately implies that 
$d(V_1^k| \sfB^{(k)}V_1^k, U_1^k) \leq o(k)$, which using the rank property for the RID gives  
\begin{align}
m_b(k) \geq d(\sfB^{(k)}V_1^k|U_1^k) \geq d(V_1^k|U_1^k)-o(k),
\end{align}
which implies the desired result $\rho_b \geq d(V_1|U_1)$. The other inequality $\rho_a \geq d(U_1|V_1)$ follows similarly. 
\end{proof}
Theorem \ref{rid_pres_prop_multi} can be extended in a natural way to the $d$-dim sources, where it can be shown that for a family of matrices to be RID-preserving for the $d$-dim source, it is necessary that their measurement rate belong to the $d$-dim region in \eqref{rate_region_dd}. 

%Using a similar technique as in Section \ref{rid_pre_single}, we can prove that the measurement rate region 
%\eqref{rate_region_2d} for the $2$-dim source, or more generally the $d$-dim region in \eqref{rate_region_dd} for the $d$-dim source, is achievable.
%The main steps can be summarized as follows:
%\begin{itemize}
%\item Take a sufficiently large $N=2^n$ of i.i.d. realization of the $d$-dim source $\bfU=(U_1,U_2,\dots,U_d)$.
%\item Multiply the $N$-dim realization in each terminal by $\bH_N$. 
%\item Use a monotonic chain rule expansion as in \cite{bilkent2012polar} to build a $d$-dim polarizing erasure process $(I^{(1)}_n, I^{(2)}_n, \dots, I^{(d)}_n)$ whose average $\bE[(I^{(1)}_n, I^{(2)}_n, \dots, I^{(d)}_n)]$ converges to a specific point in the $d$-dim region \eqref{rate_region_dd}.
%\item In each terminal $t \in [d]$ build a submatrix of $\bH_N$ by selecting those rows of $\bH_N$ for which $I_n^{(t)} (i) \geq \epsilon ^{(N)}$ for a suitable threshold $\epsilon ^{(N)}$ for which $N\epsilon ^{(N)}=o(N)$.
%\item Use the polarization property to prove that the constructed family of matrices is RID-preserving and has the desired measurement rate as the dimension $N$ tends to infinity.
%\end{itemize}

\section{Applications in Compressed Sensing}\label{sec:application}
In Compressed Sensing, the aim is to recover a structured signal $\bfx=x_1^N$ by taking only a few number of linear measurements $\bfy=\sfA\bfx$, where 
$\bfy=y_1^m$ denotes the vector of $m$ linear measurements taken via the $m \times N$ matrix $\sfA$. If the signal $\bfx$ has a sparse representation in a basis with at most $k\ll N$ nonzero elements ($k$-sparse) and if $\sfA$ is suitably designed with respect to this basis, $\bfx$ can be recovered by taking $m \ll N$  measurements \cite{CS1,CS2,CS3,CS4}.
Fix a $\delta \in (0,1)$ and consider an $N$-dimensional signal $X_1^N \in \bR^N$ whose components are 
sampled i.i.d. from the distribution 
\begin{align}\label{sig_dist}
p_X(x)=(1-\delta) \bI_0(x) + \delta p_c(x),
\end{align}
where $\bI_0(x)$ denotes a delta measure at point $0$ and where $p_c$ is a continuous probability distribution. For a large block-length $N$, almost all the realizations $\bfx=x_1^n$ of the signal $X_1^N$ have approximately $k=N\delta$ nonzero components, thus,   a sparse signal with a sparsity ratio $\delta=\frac{k}{N}=d(X)$.

Let $N=2^n$ be a power of $2$ and let $\clS^{(N)}$ be as in \eqref{sampling_pattern_1}. Let ${\sfH^{(N)}}$ be the submatrix of ${\bH}_N$ consisting of the rows in $\clS^{(N)}$ and let ${Y_1^{m(N)}={\sfH^{(N)}} X_1^N}$ be the measurements. From the RID-preserving property of $\sfH^{(N)}$ proved in Proposition \ref{single_H_rid_pres}, we have  that 
$d(X_1^N|Y_1^{m(N)})=o(N)$. From the definition of the RID, this implies that for a sufficiently large $q$, the measurements $Y_1^{m(N)}$ capture a significant fraction of the information of the quantized signal $\quant{X_1^N}_q$.

\begin{figure*}[t]
\centering
\includegraphics{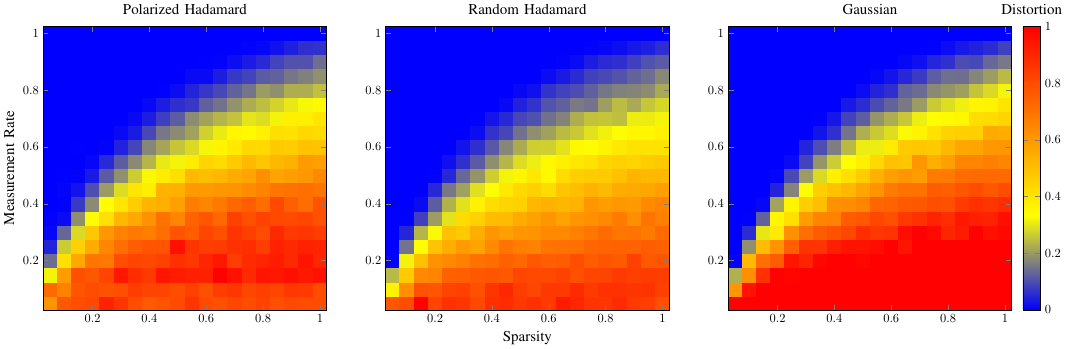}
\caption{The Rate-Distortion (RD) curve for partial Hadamard matrices and its comparison with that of random Hadamard and random Gaussian matrices. The horizontal axis indicates the sparsity level of the signal (the fraction of nonzero components in the signal), and the vertical axis shows the measurement rate (the number of measurements per dimension of the signal).}
\label{rd_comp}
\end{figure*}

In this paper, we mainly focused on the polarization of the RID as an information measure. It is interesting to know whether the RID polarization for the infinite-alphabet signals proved in this paper, can be exploited as in the case of discrete polarization for finite-alphabet sources \cite{polar_source} to build a decoder that recovers the initial signal $X_1^N$ from the collection of linear measurements $Y_1^{m(N)}$ up to a negligible distortion (e.g., error probability or $l_2$-distortion). In this section, we propose an approach to establish such an \textit{operational} aspect of the problem although we do not  prove it.

Let $\epsilon_N=2^{-N^\beta}$ for some $\beta \in (0,\frac{1}{2})$ be the threshold value used for constructing $\sfH_N$  in \eqref{sampling_pattern_1}. Then, for a sufficiently large block-length $N$, it results that
\begin{align}\label{h_rel_op}
\lim_{q \to \infty} \frac{H(\quant{X_1^N}_q|Y_1^{m(N)})}{\log_2(q)}=d(X_1^N|Y_1^{m(N)}) \leq N \epsilon_N.
\end{align}
Let $\alpha_N \in \bR_+$ and $q_N\in \bN$ be two sequences of $N \in \{2^n: n\in \bZ_+\}$ such that $\lim_{N\to \infty} q_N=\infty$ and
\begin{align}\label{h_rel_op2}
H(\quant{X_1^N}_{q_N}|Y_1^{m(N)})\leq \alpha_N N\epsilon_N \log_2(q_N)=:p_N.
\end{align}
Note that $\alpha_N$ is a scaling factor used to ensure that a sequence of $q_N$  satisfying \eqref{h_rel_op} exists.
We prove that under the stated conditions, if $p_N$ tends to zero as $N$ tends to infinity, then we can decode the quantized signal $\quant{X_1^N}_{q_N}$ with a vanishing error probability $p_N$, using the \textit{MAP} (Maximum a posteriori Probability) decoder. We use the following simple lemma. 
\begin{lemma}\label{lemma_op}
Let $D$ be a discrete random variable taking values in the countable alphabet $\clD$ and let $Y$ be an arbitrary random variable, jointly distributed with $D$, such that the conditional distribution (probability mass function) $p(d|y)$ is well-defined almost surely. Let $\clE=\{(d,y): \widehat{D}(y) \neq d\}$ be the error event of the MAP decoder  defined by $\widehat{D}(y)=\argmax_{d'\in \clD} p(d'|y)$. Then, the average error probability satisfies $\bP[\clE]\leq H(D|Y) \log_2(e)$, where $H(D|Y)$ denotes the conditional entropy of $D$ given $Y$ in bits.\hfill {\LARGE $\square$}
\end{lemma}
\begin{proof}
Proof in Appendix \ref{lemma_op_app}. 
\end{proof}
\noindent Using Lemma \ref{lemma_op}, we can see from \eqref{h_rel_op2} that the quantized components $\quant{X_1^N}_{q_N}$ can be recovered up to an average error probability  $p_N\log_2(e)$. This implies that, with a very high probability, the desired signal $X_1^N$ can be recovered up to a vanishing distortion $\frac{1}{q_N}$ provided that $p_N$ tends to $0$. We state this as the following conjecture.
\begin{conjecture}\label{pN_conj}
Let $N \in \{\interv{2^n}{n\in \bZ_+}\}$ and let $\beta\in (0,\frac{1}{2})$ and  $\epsilon_N=2^{-N^\beta}$  as before. There exists a scaling factor $\alpha_N$ and a quantization factor $q_N \stackrel{N\to \infty}{\longrightarrow} \infty$ with $p_N:=\alpha_N N\epsilon_N \log_2(q_N)\stackrel{N\to \infty}{\longrightarrow}0$. 
\end{conjecture}
For example, if we set $\alpha_N=1$ for all $N$, due to the logarithmic dependence of $p_N$ on $q_N$, a sequence $q_N=o(2^{2^{N^\beta}})$ would be sufficient for the Conjecture \ref{pN_conj} to be true. Considering the doubly-exponential growth rate of $2^{2^{N^\beta}}$ as a function of $N$, we believe that such a sequence $q_N$ should exist. This would establish the operational performance of our constructed matrices for Compressed Sensing of i.i.d. sources. Although we do not directly prove Conjecture \ref{pN_conj}, we use numerical simulations in Section \ref{sec:sim} to illustrate that our constructed polarized Hadamard matrices along with the off-the-shelf low-complexity $l_1$-norm minimization algorithm in Compressed Sensing (instead of the more complicated MAP decoder) still have a promising operational performance. 

Using partial Hadamard matrices has several practical advantages. Their components are $\pm 1$ and can be robustly implemented as on-off pattern in many practical measurement devices and easily stored in a computer.  Partial Hadamard matrices also yield computationally efficient recovery algorithms. In brief, a crucial step in all recovery algorithms in Compressed Sensing is computing  $\sfA^\transp \bfy$ (\textit{matched-filtering}), in which the inner product of the columns of the measurement matrix $\sfA$ with the observations $\bfy=\sfA\, \bfx$ is calculated. Using the structure of the Hadamard matrices, this can be done with $O(N \log_2(N))$  rather than $O(N^2)$ operations needed for the traditional matrix-vector multiplication. Even for small dimensions such as $N=1000$ this is around $100$ times faster.

\subsection{Simulation Results}\label{sec:sim}
In this section, we assess the operational performance of the partial Hadamard matrices constructed in Section \ref{rid_pre_single} via numerical simulations. 

\subsubsection{Measurement Matrix and Recovery Algorithm}
For simulations, we use a zero-mean and unit-variance sparse distribution as in \eqref{sig_dist}
\begin{align}\label{sig_dist2}
p_X(x)=(1-\delta) \bI_0(x) + \delta p_c(x), 
\end{align}
where $\bI_0(x)$ denotes the delta measure at zero, and where $p_c$ is a fixed zero-mean continuous distribution with a variance $\frac{1}{\delta}$. We do the simulations for  different sparsity levels $\delta \in \{0.0,0.1,\dots,0.9,1.0\}$ of the underlying signal. Note that for a given $\delta$ in this list, the RID of the generated signal $X\sim p_X$ is given by $d(X)=\delta$. We use the mean square error (MSE) as the distortion measure $d(\bfx, \widehat{\bfx})=\|\bfx -\widehat{\bfx}\|_2^2$ between the target signal $\bfx$ and the estimate $\widehat{\bfx}$ obtained via the recovery algorithm.  The simulations are done with the Hadamard matrix of order $N=1024$. 
To build the measurement matrix $\sfA$, we sort the rows of  ${\bH}_N$ according to their conditional RIDs and select those rows with highest RID.  We use the $l_1$-norm minimization algorithm to recover the signal:
\begin{align}
\widehat{\bfx}(\bfy)=\argmin_{\bfw \in \bR^N} \|\bfw\|_1 \text{ \ subject to \ } \bfy=\sfA\bfw, \label{ell1_alg}
\end{align}
where the input to this algorithm is the vector of linear measurements $\bfy=\sfA\, \bfx$ for the given signal $\bfx$. We use the CVX package \cite{boyd2004convex} to solve \eqref{ell1_alg}.

\subsubsection{Comparison with other Measurement Matrices}
We compare the performance of our constructed matrices with random Hadamard matrices and random Gaussian matrices extensively used in Compressed Sensing. Fig.\,\ref{rd_comp} illustrates the resulting rate-distortion curve of the $l_1$-norm minimization \eqref{ell1_alg}  for these three families of matrices. It is seen that our constructed matrices have a performance very close to that of other two families, while being deterministically constructed.

%\section{Acknowledgement}
%We would like to thank the anonymous reviewers for their valuable comments. E.A. was  partly supported by NSF grant CIF-1706648.

\section{Conclusion}
In this paper, we generalized the definition of RID as an information measure from scalar random variables initially proposed in \cite{renyi} to a larger family of vector-valued random variables. We proved that for such a family the joint and the conditional RIDs are well-defined and can be computed with a closed form formula. Using this, we proved that the RID of a sequence of i.i.d. nonsingular random variables polarizes to the extreme values of $0$ and $1$ when transformed by Hadamard matrices. We also gave a closed-form expression for the polarization pattern using the BEC polarization in the discrete case. This gives a natural counterpart of the finite-alphabet source polarization in the infinite-alphabet case. We investigated some of the applications of the new polarization phenomenon in Compressed Sensing.

\appendices
\section{Proof of Proposition \ref{prop:res_prop}}\label{prop:res_prop_app}
For the first part, let $\scrR_1=\clR_1$, and for $i=2,3,\dots, p$, set $\scrR_i=\cup_{\ell=1}^i \clR_\ell$. Note that since $\{\clR_i\}_{i=1}^p$ is a partition of the rows of $\sfA$, we have $\scrR_p=[m_a]$. Using the definition of $\res$ in \eqref{eq:res_def}, we obtain  
\begin{align}
\sum_{i=1}^p &\res\Big[\sfA[\clR_i]\big|\sfA[\cup_{\ell=1}^{i-1} \clR_\ell]; \clC\Big ]= \rank\big[\sfA[\scrR_1]_\clC\big] \\
&+ \sum_{i=2}^p \Big (\rank\big[\sfA[\scrR_i]_\clC\big]-\rank\big[\sfA[\scrR_{i-1}]_\clC\big] \Big )\\
&=\rank\big[\sfA[\scrR_p]_\clC\big]=\rank\big[\sfA_\clC\big].
\end{align} 
To prove the rank-1 innovation property in the second part, first note that from the definition of $\res$ operator in \eqref{eq:res_def}, $\res\big[\sfa,\check{\sfa}|[\sfA, \check{\sfA}] ; \clC \sqcup \check{\clC} \big ] \in \{0,1\}$. In particular, it is zero if adding the individual row $\big[\sfa, \check{\sfa}\big]$ does not increase the rank of the matrix $[\sfA, \check{\sfA}]$ at column set $\clC \sqcup \check{\clC}$, where in that case, by simply restricting to $\clC$ or $\check{\clC}$, we must have $\res[\sfa|\sfA; \clC]=\res[\check{\sfa}|\check{\sfA}; \check{\clC}]=0$. This immediately gives  the desired inequality in \eqref{rank1_innov}. Moreover, if $\sfA$ and $\check{\sfA}$ have non-overlapping set of nonzero rows, $\res\big[\sfa,\check{\sfa}|[\sfA, \check{\sfA}] ; \clC \sqcup \check{\clC} \big ]$ would be $1$ if and only if either $\sfa$ increases the  rank of $\sfA$ at column set $\clC$, or $\check{\sfa}$ increases the rank of $\check{\sfA}$ at column set $\check{\clC}$, or both, where in that case the reverse inequality in \eqref{rank1_innov} also holds. This completes the proof.  

\section{Proof of Theorem \ref{RID_maintheorem}}\label{RID_maintheorem_app} 
To simplify the proof, we first prove following two lemmas.
\begin{lemma}\label{lem:disc_useful}
Let $\bfX$ and $\bfY$ be random vectors in $\clL$. Suppose that there is a matrix $\sfL$ such that  $\bfY=\sfL \bfX$. Then
\begin{align}\label{eq:disc_useful}
\lim_{q\to \infty} \frac{H(\quant{\bfY}_q|\quant{\bfX}_q)}{\log_2(q)}=0.
\end{align}
\end{lemma}
\begin{proof}
For a vector $\bfV$, we define $\Delta[\bfV]=\bfV-\quant{\bfV}_q$ as the vector of quantization residual. Let $\Deltam_1=\Delta[\bfX]$, and $\Deltam_2=\Delta[\sfL\quant{\bfX}_q]$. Then, we have\begin{align*}
\quant{\bfY}_q&=\quant{\sfL \bfX}_q=\quant{\sfL\quant{\bfX}_q+ \sfL\Deltam_1}_q\\
&=\quant{\quant{\sfL\quant{\bfX}_q}_q +\Deltam_2 + \sfL\Deltam_1 }_q\\
&\stackrel{(i)}{=}\quant{\sfL\quant{\bfX}_q}_q +\quant{\Deltam_2 + \sfL\Deltam_1 }_q,
\end{align*}
where in $(i)$ we used the fact that $\quant{\sfL\quant{\bfX}_q}_q$ is already in the quantized form, so it can go outside  the quantization operator. 
Since $\quant{\sfL\quant{\bfX}_q}_q$ is a function of $\quant{\bfX}_q$, we obtain that
\begin{align}
H(\quant{\bfY}_q&|\quant{\bfX}_q)=H(\quant{\Deltam_2 + \sfL\Deltam_1}_q|\quant{\bfX}_q).\label{h_bound_lemma}
\end{align}
Let $\boldsymbol{\Xi}=\Deltam_2 + \sfL\Deltam_1$. Applying the triangle inequality we have  
\begin{align}
\|\boldsymbol{\Xi}\|_\infty& \leq \|\Deltam_2\|_\infty + \|\sfL\Deltam_1\|_\infty \leq \frac{1}{q} (1+ \|\sfL\|_{\infty,\infty} ),
\end{align}
where $\|\sfL\|_{\infty, \infty}= \sup_{\bfx: \bfx\neq 0} \frac{\|\sfL \bfx\|_\infty}{\|\bfx\|_\infty}$ is the operator norm of $\sfL$, and where $\frac{1}{q}$ results from the fact that every component of the  quantization residual is always bounded by $\frac{1}{q}$. This implies that $\quant{\boldsymbol{\Xi}}_q$ can take at most $\ceil{1+ \|\sfL\|_{\infty,\infty}}^n$ different values, thus, from \eqref{h_bound_lemma}, we have that $H(\quant{\bfY}_q|\quant{\bfX}_q)$ is upper bounded, independent of the value of $q$, by $n \log_2(\ceil{1+ \|\sfL\|_{\infty, \infty}})$, where $n$ denotes the dimension of the vector $\bfY$, and where for a real number $r$, we define $\ceil{r}=\min\{n\in \bZ: n\geq r\}$. Hence, dividing \eqref{h_bound_lemma} by $\log_2(q)$ and taking the limit as $q$ tends to infinity, we obtain the desired result. 
\end{proof}

\begin{remark}\label{rem:disc_useful}
Note that Lemma \ref{lem:disc_useful} still holds if we replace the quantized values $\quant{\bfX}_q$ in \eqref{eq:disc_useful} by the unquantized random variables $\bfX$, or keep any mixture thereof. \hfill $\Diamond$
\end{remark}

\begin{lemma}\label{lem:disc_useful2}
Suppose that all the conditions of Lemma \ref{lem:disc_useful} hold, and let $\bfZ$ be another vector in $\clL$ with the same dimension as $\bfY$. Then
\begin{align*}
\limsup_{q\to \infty} \frac{H(\quant{\bfZ+\bfY}_q|\quant{\bfX}_q)}{\log_2(q)}=\limsup_{q\to \infty} \frac{H(\quant{\bfZ}_q|\quant{\bfX}_q)}{\log_2(q)},
\end{align*}
with a similar equality holding for $\liminf$ instead of $\limsup$.
\end{lemma}
\begin{proof}
The proof follows from Lemma \ref{lem:disc_useful}. We have
\begin{align}\label{dis_useful_form}
H(&\quant{\bfZ+\bfY}_q|\quant{\bfX}_q)=H(\quant{\bfZ+\bfY}_q|\quant{\bfX}_q,\quant{\bfZ}_q)\nonumber\\
&+H(\quant{\bfZ}_q|\quant{\bfX}_q)- H(\quant{\bfZ}_q|\quant{\bfZ+\bfY}_q, \quant{\bfX}_q).
\end{align}
Note that $\bfY$ is a linear function of $\bfX$, thus, there are matrices $\sfL_1$ and $\sfL_2$ such that $\bfZ+\bfY=\sfL_1 [\bfX;\bfZ]$ and $\bfZ=\sfL_2[\bfZ+\bfY;\bfX]$. Dividing both sides of \eqref{dis_useful_form} by $\log_2(q)$ and using Lemma \ref{lem:disc_useful} yields that the first and the last term on the right hand side of \eqref{dis_useful_form} tend to $0$ as $q$ tends to infinity. Thus, applying $\limsup$ and $\liminf$, we obtain the desired result. 
\end{proof}

 Using Lemma \ref{lem:disc_useful2} and Remark \ref{rem:disc_useful}, we now prove Theorem \ref{RID_maintheorem}. For the first part, from the definition of  RID in \eqref{rid_def}, we have
\begin{align}
H(\quant{X_1^n}_q)&\doteq H(\quant{X_1^n}_q|\clC)\label{RID_maintheorem_4}\\
&\doteq H(\quant{\sfA_\clC \bfC_\clC + \sfA_{\clC^c} \bfD_{\clC^c}}_q|\clC)\label{RID_maintheorem_1}\\
&\doteq H(\quant{\sfA_\clC \bfC_\clC + \sfA_{\clC^c} \bfD_{\clC^c}}_q|\clC, \bfD_{\clC^c})\label{RID_maintheorem_2}\\
&\doteq H(\quant{\sfA_\clC \bfC_\clC}_q|\clC, \bfD_{\clC^c})\label{RID_maintheorem_3}\\
&=\bE_{\clC} \Big \{ H(\quant{\sfA_\clC \bfC_\clC}_q|\clC=\clC^*, \bfD_{\clC^c} )\Big \} \label{exp_result}\\
&=\bE_{\clC} \Big \{ H(\quant{\sfA_\clC \bfC_\clC}_q|\clC=\clC^*)\Big \} \label{exp_result2}
\end{align}
where $\clC$ is the support set defined by \eqref{eq:supp_set} and $\bfC=C_1^k$ and $\bfD=D_1^k$ are the vector of continuous and discrete parts of $Z_1^k$ as defined in Section \ref{sec:RID_prop},  and where in \eqref{RID_maintheorem_4}, \eqref{RID_maintheorem_2}, and \eqref{RID_maintheorem_3}, we use the fact that $\clC$ and $\bfD_{\clC^c}$ are discrete variables with a finite entropy independent of $q$ and they can be arbitrarily added or removed from the conditioning part of the entropy. Recall that from our notation in Section \ref{notation}, $f_q\doteq h_q$ for two sequences $f,h$ (parametrized with $q\in \bN$), whenever $\limsup_{q\to \infty} \frac{f_q - h_q}{\log_2(q)}=0$. For \eqref{RID_maintheorem_3}, we used Lemma \ref{lem:disc_useful2} to remove the variable $\sfA_{\clC^c} \bfD_{\clC^c}$ in \eqref{RID_maintheorem_2} since it is a linear function of $\bfD_{\clC^c}$ appearing in the conditioning part. For  \eqref{exp_result2}, we used the independence of $\bfD_{\clC^c}$ from $\sfA_\clC \bfC_\clC$ conditioned on $\clC=\clC^*$. 

Now, consider a specific realization of the support set $\clC=\clC^*$ of size $k^*=|\clC^*|$ and notice that since $\clC$ is independent of the continuous component $\bfC$, conditioning on $\clC=\clC^*$ does not change the distribution of $\bfC_{\clC^*}$, which is a $k^*$-dimensional continuous distribution. Let $m^*=\rank(\sfA_{\clC^*})$ and let $\check{\sfA}$ be a maximal submatrix of $\sfA_{\clC^*}$ consisting of linearly independent rows of $\sfA_{\clC^*}$. Then, we have
\begin{align}
H(\quant{\sfA_{\clC^*}&\bfC_{\clC^*}}_q)=H(\quant{\sfA_{\clC^*}\bfC_{\clC^*}}_q, \quant{\check{\sfA} \bfC_{\clC^*}}_q)\\
&=H(\quant{\check{\sfA} \bfC_{\clC^*}}_q)+H(\quant{\sfA_{\clC^*}\bfC_{\clC^*}}_q|\quant{\check{\sfA} \bfC_{\clC^*}}_q).\label{second_term}
\end{align}
Note that the $\check{\sfA} \bfC_{\clC^*}$ has a well-defined $m^*$-dimensional continuous distribution, thus, $\lim_{q\to \infty} \frac{H(\quant{\check{\sfA} \bfC_{\clC^*}}_q)}{\log_2(q)}=m^*$. Moreover, from Lemma \ref{lem:disc_useful2}, the second term in \eqref{second_term} vanishes in the limit when divided by $\log_2(q)$ because $\sfA_{\clC^*}\bfC_{\clC^*}$ is a linear function of $\check{\sfA} \bfC_{\clC^*}$.
Thus, from \eqref{exp_result2}, we obtain
\begin{align}
d(X_1^n)&=\lim_{q\to \infty} \frac{H(\quant{X_1^n}_q)}{\log_2(q)}\\
&= \lim_{q\to \infty} \frac{\bE_{\clC} \Big [ H(\quant{\sfA_\clC \bfC_\clC}_q|\clC=\clC^* )\Big ]}{\log_2(q)}\label{lim_exchange}\\
&= \bE_{\clC} \Big [ \lim_{q\to \infty} \frac{ H(\quant{\sfA_\clC \bfC_\clC}_q|\clC=\clC^* )}{\log_2(q)}\Big ]\\
&=\bE_\clC \Big [ \rank(\sfA_\clC)\Big ],
\end{align}
where in \eqref{lim_exchange}, we used the fact that $\clC$ takes only finitely many values and exchanged the expectation and the limit. This completes the proof of the first part of the theorem.

To prove the second part, recall that $X_1^n=\sfA Z_1^k$ and $Y_1^m=\sfB Z_1^k$. We follow similar steps as in the first part, where from \eqref{exp_result2}, we essentially need to compute 
the expression 
\begin{align}
H(\quant{\sfA_\clC \bfC_\clC}_q|\sfB_\clC \bfC_\clC, \clC=\clC^*)
\end{align}
for different realizations of support set $\clC^*$.  Note that $\sfB_{\clC^*}$ is not necessarily full-rank. Let $\check{\sfB}$ be a maximal submatrix of $\sfB_{\clC^*}$ consisting of those rows that are linearly independent. Also, let $\check{\sfA}$ be the maximal submatrix of $\sfA_{\clC^*}$ consisting of those linearly independent rows that are also linearly independent of the rows of $\check{\sfB}$. From the definition of $\res$ operator in \eqref{eq:res_def}, it is not difficult to check that the number of rows of $\check{\sfA}$ is given by $\res[\sfA|\sfB;\clC^*]$. Hence, we have
\begin{align}
H(&\quant{\sfA_{\clC^*} \bfC_{\clC^*}}_q|\sfB_{\clC^*} \bfC_{\clC^*})=H(\quant{\sfA_{\clC^*} \bfC_{\clC^*}}_q, \quant{\check{\sfA} \bfC_{\clC^*}}_q |\check{\sfB}\bfC_{\clC^*})\nonumber\\
&=H(\quant{\check{\sfA} \bfC_{\clC^*}}_q |\check{\sfB}\bfC_{\clC^*})\label{term1}\\
&+H(\quant{\sfA_{\clC^*} \bfC_{\clC^*}}_q| \quant{\check{\sfA} \bfC_{\clC^*}}_q ,\check{\sfB} \bfC_{\clC^*})\label{term2}.
\end{align}
Note that $[\check{\sfA};\check{\sfB}]$ is a full rank matrix, thus, $[\check{\sfA} \bfC_{\clC^*};\check{\sfB}\bfC_{\clC^*}]$ has a well-defined continuous distribution. In particular, for almost all realizations of $\check{\sfB}\bfC_{\clC^*}$, the random variable $\check{\sfA}\bfC_{\clC^*}$ has a continuous distribution, thus, for the first term in \eqref{term1} we have that $\lim_{q\to \infty} \frac{H(\quant{\check{\sfA} \bfC_{\clC^*}}_q |\check{\sfB}\bfC_{\clC^*})}{\log_2(q)}=\dim(\check{\sfA} \bfC_{\clC^*})$, which is equal to $\res[\sfA|\sfB;\clC^*]$. 
For the second term in \eqref{term2}, we have that $\lim_{q\to \infty} \frac{H(\quant{\sfA_{\clC^*} \bfC_{\clC^*}}_q| \quant{\check{\sfA} \bfC_{\clC^*}}_q ,\check{\sfB} \bfC_{\clC^*})}{\log_2(q)}=0$ since $\sfA_{\clC^*} \bfC_{\clC^*}$ is a linear function of $\check{\sfA} \bfC_{\clC^*}$ and $\check{\sfB} \bfC_{\clC^*}$  appearing in the conditioning part, and the result follows from Lemma \ref{lem:disc_useful} and Remark \ref{rem:disc_useful}. Thus, we have
\begin{align}
\lim_{q\to \infty} \frac{H(\quant{\sfA_\clC \bfC_\clC}_q|\sfB_\clC \bfC_\clC, \clC=\clC^*)
}{\log_2(q)}=\res[\sfA|\sfB;\clC^*],
\end{align}
and taking the average over $\clC$, we obtain the desired result.

\section{Proof of Theorem \ref{RID_extensions}}\label{RID_extensions_app}
We use the rank characterization of the RID proved in Theorem \ref{RID_maintheorem}. The positivity simply follows from the positivity of the rank of a matrix. Also, if $d(X_1^n)=0$, from the property of the rank and the definition of $X_1^n$ in \eqref{l_def}, we have that $0 \leq d(X_i) \leq d(X_1^n)=0$ for every $i \in [n]$, which implies that all the $X_i$, $i \in [n]$, are discrete variables. The invariance results from the fact that for any $n \times k$ matrix $\sfA$, any invertible $n\times n$ matrix $\sfL$, and any subset $\clC \subseteq [k]$ of columns of $\sfA$, we have $\rank(\sfA_\clC)=\rank(\sfL \sfA_\clC)$. 
The chain rule follows from the chain rule for the $\res$ operator in Proposition \ref{prop:res_prop}: 
\begin{align}
d(X_1^n,Y_1^m)&=\bE\big[\res\big[[\sfA;\sfB];\clC\big]\big]\\
&=\bE\big[\res[\sfA;\clC]\big] + \bE\big[\res[\sfB|\sfA;\clC]\big]\\
&=d(X_1^n)+d(Y_1^m|X_1^n).
\end{align}
To prove the symmetry, note that $\mri{X_1^n;Y_1^m}=d(X_1^n)-d(X_1^n|Y_1^m)$. Thus, using the chain rule property, we obtain that $\mri{X_1^n;Y_1^m}=d(X_1^n)+d(Y_1^m)-d(X_1^n,Y_1^m)$, and the symmetry follows from the  symmetry of $d(X_1^n,Y_1^m)$. 

Finally, for the last part note that from the definition of the $\res$ operator, it results that $\rank(\sfA_\clC)=\res[\sfA|\emptyset;\clC]\geq \res[\sfA|\sfB;\clC]$. Taking the average over $\clC$, we have $d(X_1^n)\geq d(X_1^n|Y_1^m)$, which implies the desired result.

\section{Proof of Theorem \ref{multi_RID_polarization}}\label{multi_RID_polarization_app}
Note that $U_1^N$ are i.i.d. random variables obtained via a linear transform of the variables in $\clL$, thus, they belong to $\clL$. Hence, from Theorem \ref{single_RID_polarization}, it immediately results that $I_n$ is an erasure process with initial value $I_0(1)=d(U_1)$ polarizing to $\{0,1\}$. 

To prove that $J_n$ is also a polarizing erasure process, first note that the recursive structure in \eqref{had_recursive} remains intact if we transform $\sfh_i$ into $\sfh_i \otimes \sfb$ and $\bH_N$ into $\bH_N \otimes \sfb$, where $\otimes$ denotes the Kronecker product. Moreover, it is not difficult to see that there are indeed two main ingredients in the proof of Theorem \ref{single_RID_polarization}: Applying the recursive structure of $\bH_N$ as in \eqref{rec_rank_1}, in order to obtain an expression for the plus-branch ($I_n(i) \to I_n(i)^+$), and using the chain rule in \eqref{ch_rule_single}, in order to compute the minus-branch ($I_n(i) \to I_n(i)^-$). It is not difficult to check that  both conditions remain valid after the mentioned transformation. This implies that $J_n$ is also an erasure process, whose initial value, from chain rule, is given by $J_0(1)=d(V_1|U_1)$.

\section{Proof of Lemma \ref{lemma_op}}\label{lemma_op_app}
Suppose that given $Y=y$, we use the MAP decoder defined by $\widehat{D}(y)=\argmax_{d'\in \clD} p(d'|y)$ to decode $D$. Then, for any arbitrary $d\in \clD$, we have
\begin{align}
\bP[\clE|Y=y]&=1-\max_{d'\in \clD} p(d'|y)\\
&\leq 1- p(d|y)= 1- e^{\log(p(d|y))}.
\end{align}
Taking the average over the joint distribution of $(D,Y)$ and using the Jensen's inequality \cite{jensen1906fonctions} for the convex function $u \mapsto e^{u}$, we have
\begin{align}\label{eqvi_eq}
\bP[\clE]&\leq 1- e^{\bE_{D,Y}[\log(p(d|y))]}=1-e^{-H(D|Y) \log_2(e)}\\
&\stackrel{(i)}{\leq} 1-(1-H(D|Y)\log_2(e))\\
&=H(D|Y)\log_2(e),
\end{align}
where in $(i)$ we used the inequality $e^{-u} \geq 1-u$ for $u\in \bR$.
This completes the proof.

\balance
\bibliographystyle{IEEEtran}
{\small 
\bibliography{references}}

\end{document}